# Efficient generation of spin currents by the Orbital Hall effect in pure Cu and Al and their measurement by a Ferris-wheel ferromagnetic resonance technique at the wafer level

Amit Rothschild[1,†], Nadav Am-Shalom[1,†], Nirel Bernstein[1,†], Ma'yan Meron[1,†], Tal David[1,†], Benjamin Assouline[1], Elichai Frohlich[1], Jiewen Xiao[2], Binghai Yan[2], Amir Capua[1]*

[1]Department of Applied Physics, The Hebrew University of Jerusalem, Jerusalem 9190401, Israel

[2]Department of Condensed Matter Physics, Weizmann Institute of Science, Rehovot 7610001, Israel

*e-mail: amir.capua@mail.huji.ac.il
†Equal contributors

## Abstract:

**We present a new ferromagnetic resonance (FMR) method that we term the "Ferris" FMR. It is wideband, has significantly higher sensitivity as compared to conventional FMR systems, and measures the absorption line rather than its derivative. It is based on large-amplitude modulation of the externally applied magnetic field that effectively magnifies signatures of the spin-transfer torque making its measurement possible even at the wafer-level. Using the Ferris FMR, we report on the generation of spin currents from the orbital Hall effect taking place in pure Cu and Al. To this end, we use the spin-orbit coupling of a thin Pt layer introduced at the interface that converts the orbital current to a measurable spin current. While Cu reveals a large effective spin Hall angle exceeding that of Pt, Al possesses an orbital Hall effect of opposite polarity in agreement with the theoretical predictions. Our results demonstrate additional spin- and orbit- functionality for two important metals in the semiconductor industry beyond their primary use as interconnects with all the advantages in power, scaling, and cost.**

Spin currents are the primary building block of spintronics technology. Their manipulation in practical applications poses challenges in their generation and detection. The spin Hall effect (SHE) has proven to be a well-established method for generating spin currents. Heavy metals such as Pt and W have been widely used to explore the SHE due to their large spin-orbit coupling (SOC) [1-4]. The orbital counterpart of the SHE is the orbital Hall effect (OHE). In the OHE an orbital current is generated without relying on SOC [5-9], and thus is expected to overcome the penalty of the large Gilbert losses of heavy metals. Because of the short orbital lifetimes, the OHE was not observed until recently, e.g. in Cr [10] and Ti [11], while the orbital Rashba-Edelstein effect [12] was reported in CuO [13-16].

The conversion efficiency of charge current to spin current is known as the spin Hall angle, $\theta_{SH}$. The ferromagnetic resonance (FMR) based techniques [17-24] have proven pivotal for accurately determining $\theta_{SH}$. In these measurements the generated spin-transfer torque (STT) modifies the resonance linewidth by the anti-damping torque. Two common implementations of the FMR experiment are the cavity [17,24] and stripline FMR [25,26]. While the cavity FMR benefits from high sensitivity suitable for atomically thin samples, it is narrowband. On the other hand, the stripline FMR is broadband but typically has lower sensitivity. For these reasons the spin-torque FMR (STFMR) technique [19,27-30] gained popularity for quantifying $\theta_{SH}$ in which dual AC Oersted- and spin- torque excitations produce a DC voltage that is probed electrically on a pre-patterned device [31,32].

In conventional FMR techniques [17,24-26] the sensitivity is achieved by applying a small signal modulation to the external magnetic field which results in a signal of a proportionally small amplitude that represents the differential absorption. In contrast, in this work we apply a large-amplitude modulation of the externally applied magnetic field resulting in an on-off modulation of the absorption leading to a greater sensitivity to the FMR signal. Additionally, it produces a signal proportional to the actual absorption spectrum. The modulation is achieved by placing permanent magnets on a spinning disk hence we term the technique the *Ferris FMR*. The technique is implemented in a stripline configuration making it broadband and well-suited for atomically thin films. Most importantly, the large-amplitude modulation turns out to expand the linewidth by ∼ 2 - 5 times. Consequently, $\theta_{SH}$ can be reliably resolved at lower applied currents and the measurement of $\theta_{SH}$ becomes possible at the wafer scale without patterning devices. Using the Ferris FMR we demonstrate the ability to generate a sizeable STT in pure Cu and Al as predicted by recent theory of the OHE [33,34]. To that end, we use the SOC of a thin layer of Pt that converts the orbital current to a spin current [10,14,34,35]. A higher effective $\theta_{SH}$, $\theta_{SH}^{eff}$, of the Cu based system is found as compared to Pt while at the same time the Gilbert losses are lower. Interestingly, Al displays a negative $\theta_{SH}^{eff}$ in agreement with theoretical predictions.

The Ferris FMR setup is presented in Fig. 1(a). The magnetic film is placed on a stripline waveguide at the output of which the power of the microwave signal is monitored using an RF diode and a lock-in amplifier. The externally applied magnetic field, $H(t)$, is generated using pairs of permanent magnets of opposite polarity resulting in an in-plane field required for the measurement of $\theta_{SH}$. A sinusoidally modulated profile of amplitude $H_0$ results (Fig. 1(b)) that is varied by translating the spinning disc.

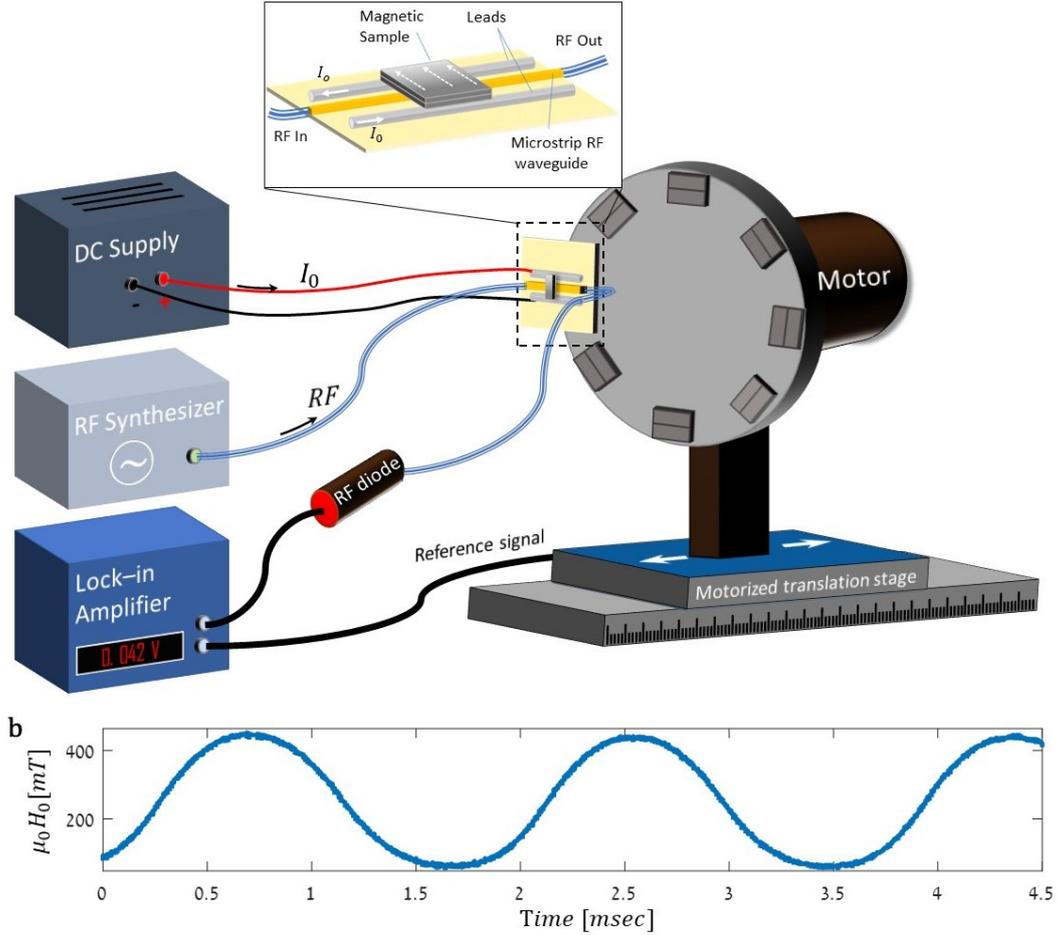

**Fig. 1. (a) Ferris FMR setup. $H(t)$ is generated by rotating a magnetic disc. After passing through the waveguide, the RF signal is detected on an RF diode detector. Left inset: two flexible leads are placed on each side of the waveguide to pass current through the sample. (b) Measured temporal profile of $H(t)$.**

The signal recorded on the lock-in amplifier, $V_{LI}(H_0)$, was determined by projecting $V_{rf}(t)$ on the fundamental harmonic. Figure 2(a) presents $V_{LI}(H_0)$ in addition to an ideal Lorentzian absorption line. $V_{LI}(H_0)$ is asymmetric, its peak occurring at $H_{peak}$, appears at a slightly higher field than $H_{res}$, and it is wider as compared to the ideal Lorentzian lineshape. We define the full width at half maximum (FWHM) linewidth expansion ratio by $A_c$. In the absence of anisotropy, we find $A_c = 2.85$ where in the general case it has to be calculated numerically as described above.

An example of the calculated and measured asymmetric spectra for a bilayer of 7.5 Pt/7.5 Py (numbers indicate layer thicknesses in $nm$) in a die of $0.75 \times 0.5\ cm^2$ ($W \times L$) are presented in Figs. 2(c) & 2(d). Typical magnet to sample distances were $22 - 13\ mm$ while the homogeneity of $H(t)$ across the sample was verified by testing smaller $\sim 0.5 \times 0.5\ mm^2$ samples for which the response remained unchanged. Each magnet was $10 \times 17\ mm^2$ placed on a disc of $13\ cm$ in diameter. The minimal incremental movement of the translation stage was $0.05\ \mu m$ so that $H_0$ was controllable to an accuracy of $0.6\ \mu T$ at the closer end, well beyond any requirement of an FMR experiment. $\omega_{mod}/2\pi$ was $500\ Hz$. The films in this work were grown by magnetron sputtering at a base pressure of $7 \times 10^{-10}\ Torr$ on Si/SiO₂ substrates and were capped with TaN ($2.5\ nm$). $f_{res}$ versus $H_{peak}$ follows Kittel's formula as seen in Fig. 2(e) leading to $M_s = 7.3 \cdot 10^5\ A/m$ which was extracted using the approximation $H_{peak} \cong H_{res}$. The results were verified using an optical STFMR (OSTFMR) experiment (open red circles) described in Ref. [36]. Figure 2(f) presents the measured FWHM linewidth, $\Delta H$, as a function of $f_{res}$ resulting in $\alpha = 0.0135 \pm 0.0002$ obtained OSTFMR measurements (red trace) resulted in a close value of $\alpha = 0.0124 \pm 0.0001$ where the difference may be related to the device versus film level dynamics. The detection limit of the Ferris FMR was estimated to be $1.55 \times 10^{11}\ \mu_B$ from a $520 \times 410\ \mu m^2$ sample having $SNR = 160.9$, that is $1 - 2$ orders of magnitude more sensitive than conventional FMR systems.

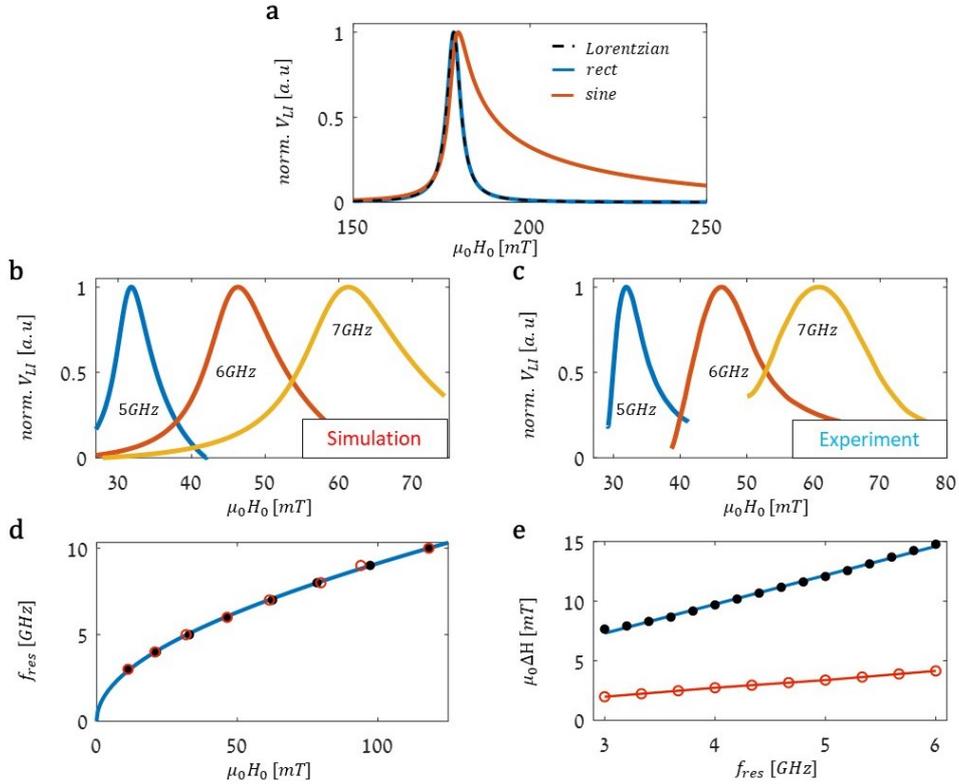

**Fig. 2. (a) Calculation of an ideal Lorentzian absorption line (dashed black), and lineshapes obtained by rectangular (solid blue) and sinusoidal modulation profiles (solid red) at $5\ GHz$ without shape anisotropy. b) Simulated spectra at 5, 6, and $7\ GHz$ with $\alpha = 0.01$. (c) Measured spectra at 5, 6, and $7\ GHz$. (d) Measured $f_{res}$**

vs. $H_{peak}$ using Ferris FMR (black circles) and OSTFMR (open red circles). Solid line indicates fitted Kittel's formula. (e) $\Delta H$ as function of $f_{res}$ measured by the Ferris FMR (black circles) and OSTFMR (open red circles). Blue solid line indicates linear fit.

To demonstrate a measurement of $\theta_{SH}$ we pass a charge current in the well-studied Pt/Py bilayer that modulates the effective $\alpha$ by the anti-damping torque in the usual manner [19]. A calculation of the absorption spectra for three different $\alpha$ values is presented in Fig. 3(a) revealing that $H_{peak}$ shifts in addition to the broadening of $\Delta H$. In the Pt/Py system $\theta_{SH}^{eff}$ represents the intrinsic $\theta_{SH}$ of Pt, $\theta_{SH}^{Pt}$, normalized by the transparency of the Pt/Py interface, $T_s$. $\theta_{SH}^{Pt}$ accounts for the spin diffusion length, $\lambda_{SD}$, and is layer thickness specific. The measurements here and in the ones that follow were carried out in dies of $0.75 \times 0.5\ cm^2$ ($W \times L$). As compared to device level measurements, achieving sufficiently high current densities, $J_c$, to drive a sizeable SHE is more difficult due to joule heating. However, in the Ferris FMR technique the linewidth is expanded by $A_c$ so that a measurable torque should already be obtained at $A_c^{-1}$ of typical applied $J_c$. To reduce the joule heating we reduce the resistance of the film by measuring a 25 Pt/5 Py bilayer that has a relatively thick Pt layer, well beyond $\lambda_{SD}$. $J_c$ was driven through two semi flexible conducting leads placed to the sides of the waveguide $\sim 2\ mm$ apart (Fig. 1(a), left inset) while the resistance was monitored to assure adequate contact. The geometrical arrangement of the experiment is illustrated in the lower inset of Fig. 1(a). The measured $J_c$ dependent $\Delta H$ modulation is presented in Fig. 3(b) and consists of a symmetric part in $J_c$, $\Delta H_S$, represented by the red solid line and an antisymmetric part, $\Delta H_A$, presented in Fig. 3(c). $\Delta H_S$ stems from joule heating [24] while $\Delta H_A$ stems from the generated STT. A sizeable antisymmetric SHE induced linewidth broadening is seen for $|J_c|$ of up to $4 \cdot 10^9\ A/m^2$ which is generally lower than typical $J_c$ applied in $\theta_{SH}$ measurements. Following Ref. [37] we extract $\theta_{SH}^{eff}$.

$\alpha = 0.0144$ was determined from the Ferris FMR measurement. Accordingly, $\theta_{SH,Pt}^{eff} = 0.09 \pm 0.01$ results, agreeing well with measured values for Pt [19,36-38]. Figures 3(d) & 3(e) present the measured symmetric and antisymmetric parts of the $J_c$ dependent $H_{peak}$, $H_{peak}^S$ and $H_{peak}^A$, respectively. $H_{peak}^S$ stems from the joule heating. However, extraction of $\theta_{SH}$ from $H_{peak}^A$ is not straight forward. From Fig. 3(a) it is seen that the influence of $\alpha$ on $H_{peak}$ is significantly smaller than its effect on $\Delta H$. In addition, $H_{peak}^A$ is also affected by the Oersted field contribution of the Pt layer which is antisymmetric in $J_c$ and masks the STT contribution.

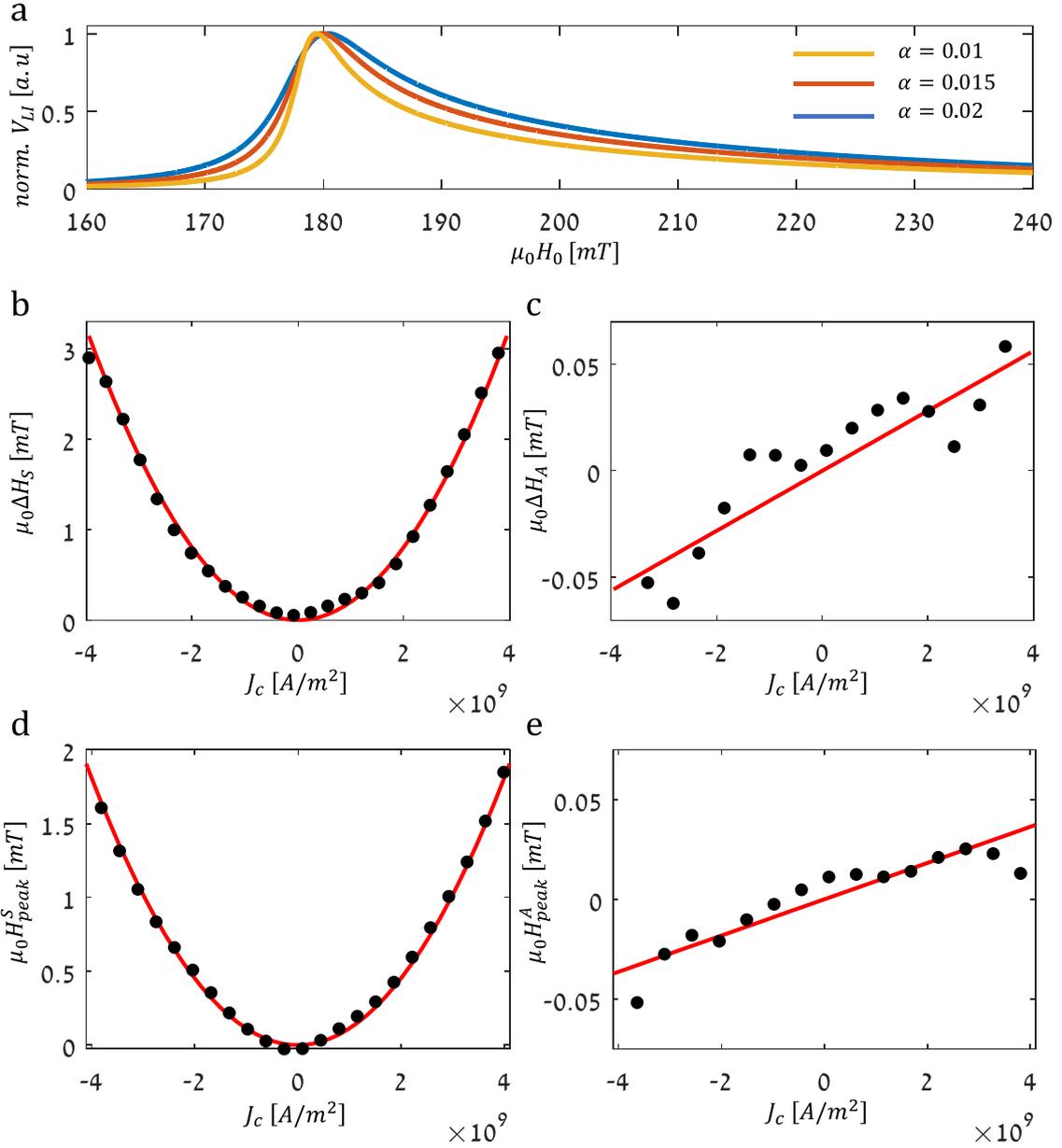

Fig. 3. SHE measurement in Pt. (a) Calculation of the Ferris FMR responses for $\alpha$ of 0.01 (yellow), 0.015 (red), and 0.02 (blue) at $5\ GHz$. (b) Measured $\Delta H$ as a function of $J_c$ (black dots) together with $\Delta H_S$ of the fit (red solid line). (c) Antisymmetric component of (b). Red solid line is a fit of $\Delta H_A$. (d) Measured $H_{peak}$ as a function of $J_c$ (black dots) and fit of $H_{peak}^S$ (red solid line). (e) Antisymmetric component of (d).

Next, we demonstrate the ability to efficiently generate STT using Cu and Al. According to OHE theory, Cu is predicted to be capable of generating significant orbital current that arises from $sd$ hybridization resulting in sizable orbital Hall conductivity that is comparable in magnitude to the spin Hall conductivity of Pt [33,34]. In contrast to spin currents, orbital currents cannot exert a torque because of the lack of exchange coupling between orbital angular momentum and the local magnetic moment.

Therefore, we introduce SOC at the Cu/Py interface by adding a thin layer of Pt that converts the orbital current into a measurable spin current [10,11,34,39]. We use the trilayer of $X_{OHE}$/1 Pt/7.5 Py where $X_{OHE}$ =13 Cu, 30 Cu, and 50 Al, as illustrated in Fig. 4(a). The high conductivity of Cu as compared to Pt enables to carry out the measurement using a thinner Cu layer. We first examine the structure of 13 Cu/1 Pt/7.5 Py. The measured $\Delta H_A$ is summarized in Fig. 4(b) as function of an effective $J_c$ that is the weighted current density average passing in the spin current generating. The measured STT stems from multiple spin and orbital dependent processes. The orbital current generated in the Cu layer is converted at the Cu/Pt interface into spin current that follows diffusive transport into the Py and transparency at the Pt/Py interface. It is well known that the Rashba-Edelstein effect in Cu/Pt interfaces is negligible [40,41] and so is the SHE of Cu. Lastly, is the contribution of the spin current generated by the SHE of Pt. Accordingly, $\theta^{eff}_{SH,Cu_{13}} = 0.110 \pm 0.012$ (see supplementary note for further detail).

Figure 4(b) also presents data for a trilayer having a thicker Cu layer of 30 Cu/1 Pt/7.5 Py. An even larger modulation of $\Delta H$ is seen leading to $\theta^{eff}_{SH,Cu_{30}} = 0.160 \pm 0.011$. The interfaces are identical to those of the $Cu_{13}$ based trilayer while $J_{c,Pt}$ reduces to $0.14 \cdot J_{c,Cu}$ so that the contribution of the SHE within the Pt conversion layer is even further diminished. Yet $\theta^{eff}_{SH,Cu_{30}} > \theta^{eff}_{SH,Cu_{13}}$ readily indicating that $J_s$ stems from the bulk of the Cu by the generation of orbital current.

The results were verified by conventional device-level measurements using the OSTFMR technique. $350 \times 450 \ \mu m^2 \ (W \times L)$ devices were fabricated from the same $Cu_{30}$ based film. The measurement was carried out applying higher $J_c^{eff}$ of up to $10^{10} \ A/m^2$ (dashed yellow line of Fig. 4(b)). $\Delta H_A$ of the OSTFMR is narrower as expected whereas $\theta^{eff}_{SH,Cu_{30}} = 0.12 \pm 0.01$, confirming the Ferris FMR measurements. However, this value is slightly lower than $\theta^{eff}_{SH,Cu_{30}}$ measured by the Ferris FMR. The difference may be attributed to the joule heating and requires further investigation. In a second test, the conversion layer was removed resulting in the bilayer of 30 Cu/7.5 Py Figure 4(b) readily shows that the modulation of $\Delta H_A$ is diminished with $\theta^{eff}_{SH,no\ Pt} = -0.0006 \pm 0.005$.

When Cu is replaced by Al, $sp$ orbital hybridization takes place. In this case an orbital current of opposite polarity was predicted [33]. The resistivity of Al is higher than Cu, therefore, our measurements were carried out on 50 Al/1 Pt/7.5 Py and are presented in Fig. 4(b). $\Delta H_A$ reveals the predicted negative $\theta_{SH}$ [33] from which we find $\theta^{eff}_{SH,Al} = -0.12 \pm 0.01$. The negative $\theta^{eff}_{SH,Al}$ was also verified using the OSTFMR device-level measurement (Fig. 4(b)) leading to $\theta^{eff}_{SH,Al} = -0.08 \pm 0.01$. Once more, this value is slightly lower than $\theta^{eff}_{SH}$ obtained from the Ferris FMR. A possible role of the Al/Pt interface may exist.

$\alpha$ critically depends on SOC. Since the OHE does not rely on SOC, it is anticipated to be capable of producing a high $\theta_{SH}$ with low $\alpha$. In the material systems at hand, $\alpha$ is enhanced by spin pumping into the adjacent nonmagnetic metal. Figure 4(c) presents the $\alpha$ measurements. Pt is well known to be an efficient sink for spin angular momentum and indeed the largest damping is found in the 25 nm Pt based bilayer with $\alpha_{Pt_{25}} = 0.0144 \pm 0.0003$. When the Pt layer is replaced by Cu, SOC is reduced and $\alpha$ decreases as seen for 30 Cu/7.5 Py resulting in $\alpha_{no\ Pt} = 0.0121 \pm 0.0002$. In this case spin diffusion into the Cu takes place. Since $\lambda_{SD} \approx 450\ nm$ in Cu, the full thickness of Cu layer contributes to the losses. When the $1\ nm$ Pt conversion layer is introduced, $\alpha$ increases only slightly to $\alpha_{Cu_{30}} = 0.0124 \pm 0.0002$ indicating that the additional losses stemming from the conversion layer are marginal and that the primary contribution remains the bulk of the Cu. As compared to the $Pt_{25}$ bilayer, the $Cu_{30}$ based trilayer displays a higher $\theta_{SH}^{eff}$ with lower $\alpha$. When the thickness of the Cu film is reduced as in the $Cu_{13}$ trilayer, $\alpha$ reduces significantly to $\alpha_{Cu_{13}} = 0.010 \pm 0.002$ providing further evidence that $\alpha$ stems from bulk of the Cu film following spin propagation through the Pt layer. Finally, Al results in $\alpha_{Al} = 0.0133 \pm 0.003$ illustrating once more that a higher $\theta_{SH}^{eff}$ is achievable with lower $\alpha$ as compared to the SHE of Pt.

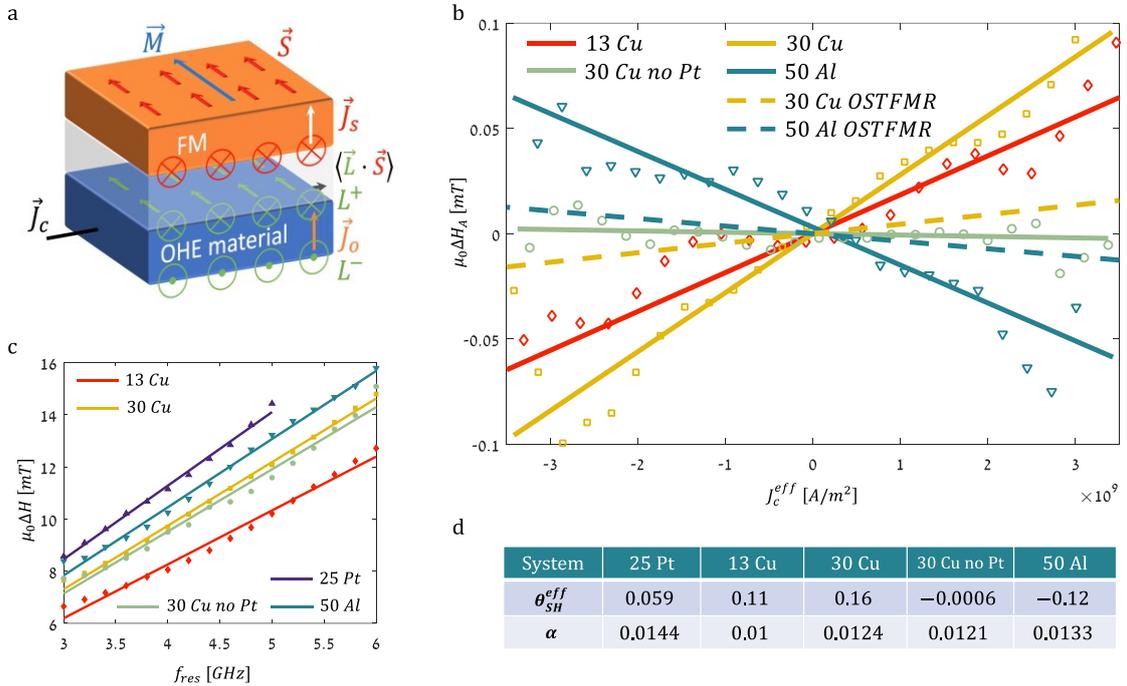

**Fig. 4. OHE measurement in Cu and Al. (a) Schematic of the trilayer system. (b) Measured $\Delta H_A$ as a function of $J_c^{eff}$ in the $Cu_{13}$ (red), $Cu_{30}$ (yellow), $Cu_{30}$ without $Pt_1$ conversion layer (light green), and $Al_{50}$ (dark green) based systems at $5\ GHz$. Solid line represents fit to measurement. Dashed lines represent the fitted data of**

the OSTFMR measurements. (c) $\Delta H$ vs. $f_{res}$. Color code same as in (b). $Pt_{25}$ data indicated in purple. Traces are shifted to cross the origin for clarity. (d) Summary of $\theta_{SH}$ and $\alpha$.

Al and Cu are key metals in the semiconductor industry that offer superior current-carrying capacity due to their high conductivity and excellent heat dissipation. Having overcome the fabrication challenges such as electromigration and Si contamination, they are currently considered the metals of choice for interconnects in high-volume applications which benefit from improved speed and power performance at attractively low-cost. Our results demonstrate additional spin- and orbit- functionality for Cu and Al beyond their use as interconnects. In agreement with OHE theory, they displayed efficient spin current generation and low $\alpha$, as compared to the SHE of Pt, while at the same time offering complementary spin logic. These observations were obtained in die-level measurements and are expected to facilitate the exploration of light metals for spin- and orbit- based technologies.